\documentclass[12pt]{article}
\usepackage{amsmath,amssymb,epsfig,youngtab}
\makeatletter \@addtoreset{equation}{section} \makeatother
\renewcommand{\theequation}{\thesection.\arabic{equation}}
\newcommand{\ba}{\begin{array}}
\newcommand{\ea}{\end{array}}
\newcommand{\beq}{\begin{equation}}
\newcommand{\eeq}{\end{equation}}
\newcommand{\bea}{\begin{eqnarray}}
\newcommand{\eea}{\end{eqnarray}}

\def\bce{\begin{center}}
\def\ece{\end{center}}

\def\nonu{\nonumber}

\def\pa{\partial}

\def\be{\beta}

\def\La{\Lambda}

\def\eps6{{\displaystyle \mathop{\epsilon}^{6}}{}}

\def\nab6{{\displaystyle \mathop{\nabla}^{6}}{}}

\def\0{{\sst{(0)}}}
\def\1{{\sst{(1)}}}
\def\2{{\sst{(2)}}}
\def\3{{\sst{(3)}}}
\def\4{{\sst{(4)}}}
\def\5{{\sst{(5)}}}
\def\6{{\sst{(6)}}}
\def\7{{\sst{(7)}}}
\def\8{{\sst{(8)}}}

\def\ba{\begin{array}}
\def\ea{\end{array}}
\def\beq{\begin{equation}}
\def\eeq{\end{equation}}
\def\be{\begin{equation}}
\def\ee{\end{equation}}

\def\Tr{\mathop{\rm Tr}}

\def\eps{\epsilon}

\def\ba{\begin{array}}
\def\ea{\end{array}}
\def\beq{\begin{equation}}
\def\eeq{\end{equation}}
\def\be{\begin{equation}}
\def\ee{\end{equation}}

\def\Tr{\mathop{\rm Tr}}

\def\eps{\epsilon}

\newcommand{\bean}{\begin{eqnarray*}}
\newcommand{\eean}{\end{eqnarray*}}

\begin{document}
\thispagestyle{empty} \addtocounter{page}{-1}
   \begin{flushright}
\end{flushright}

\vspace*{1.3cm}

 \centerline{ \Large \bf  Squashing Gravity Dual of } 
\vspace{.3cm} 
\centerline{ \Large \bf  
${\cal N}=6$ Superconformal Chern-Simons Gauge Theory   } 
\vspace*{1.5cm}
\centerline{{\bf Changhyun Ahn }
} 
\vspace*{1.0cm} 
\centerline{\it  
Department of Physics, Kyungpook National University, Taegu
702-701, Korea} 
\vspace*{0.8cm} 
\centerline{\tt ahn@knu.ac.kr
} 
\vskip2cm

\centerline{\bf Abstract}
\vspace*{0.5cm}

Four-dimensional field equations are determined for perturbations of
the quotient seven-sphere size and squashing parameter in eleven-dimensional 
supergravity. The quotient seven-sphere   
is a ${\bf S}^1$-bundle over the
${\bf CP}^3$ which is regarded as a ${\bf S}^2$-fibration 
over the base ${\bf S}^4$.
By analyzing the $AdS_4$ supergravity scalar potential, 
the holographic supersymmetric(or nonsupersymmetric) 
renormalization group(RG) flow 
from ${\cal N}=1$(or 
${\cal N}=0$) $SO(5) \times U(1)$-invariant UV fixed
point to ${\cal N}=6$(or ${\cal N}=0$) 
$SU(4)_R \times U(1)$-invariant IR fixed point 
is obtained. The three-dimensional boundary theories are described by 
superconformal Chern-Simons matter theories and a dual operator
corresponding to this RG flow is described.  

\baselineskip=18pt
\newpage
\renewcommand{\theequation}
{\arabic{section}\mbox{.}\arabic{equation}}

\section{Introduction}

The three-dimensional Chern-Simons matter theory
with gauge group $U(N) \times U(N)$ at level $k$ 
which has ${\cal N}=6$
superconformal symmetry is constructed recently in \cite{ABJM}.
They describe this gauge theory 
as the low energy limit of $N$ M2-branes probing 
${\bf C}^4/{\bf Z}_k$ singularity.
At large $N$-limit, this theory is dual to the M-theory on 
$AdS_4 \times {\bf S}^7/{\bf Z}_k$. 
In order to perform the ${\bf Z}_k$-quotient 
explicitly \footnote{
There exists a variety of orbifolds with free or nonfree
actions on ${\bf S}^7$ leading to different amount of supersymmetry 
\cite{MP}. Let us consider
M2 branes at ${\bf C}^4/\Gamma$ singularity where
 the group $\Gamma$ is generated by $\mbox{diag}( 
 e^{\frac{2\pi i}{k}}, e^{-\frac{2 \pi i}{k}}, e^{\frac{2 \pi i a}{k}},
e^{-\frac{2\pi i a}{k}})$ for some relatively prime intergers $a$ and $k$.
If $a=1, k=2$ we get maximal case ${\cal N}=8$ with horizon manifold
${\bf RP}^7$.
For $a=\pm 1, k > 2$, one gets ${\cal N}=6$ 
theory \cite{Halyo} where the corresponding field theory duals are present.
When $a \neq \pm 1$, the theory has ${\cal N}=4$ supersymmetry.
Similar ${\cal N}=4$ theory \cite{FKPZ} can be obtained from the different
orbifold ${\bf C}^2/\Gamma \times {\bf C}^2/\Gamma'$(See also \cite{Gomis}).
The orbifold ${\bf C} \times {\bf C}^3/\Gamma$ provides the ${\cal
  N}=2$ theory \cite{AOT}.
When $\Gamma$ is a binary dihedral group where ${\bf C}^2/\Gamma$ has
a $D$ type singularity and we embed
$\Gamma$ into $SU(2) \times SU(2)$, we get ${\cal N}=5$ theory.
}, 
it is natural to write
the seven-sphere ${\bf S}^7$ metric as an ${\bf S}^1$-fibration over ${\bf
CP}^3$ \cite{NP}.
When $N=2$ with $k=1,2$, the theory is equivalent to Bagger-Lambert 
theory \cite{BL,BL1,BL2}. 

It is known in \cite{AR} that 
deformation from ``round'' seven-sphere ${\bf S}^7$ to ``squashed'' one 
$\widetilde{{\bf S}}^7$ \cite{Jensen,BK}(See also \cite{ADP,DNP}) 
can be interpreted
as
renormalization group(RG) flow from UV fixed point to IR fixed point,
via AdS/CFT correspondence \cite{Maldacena,Witten,GKP}, by
analyzing the four-dimensional effective Lagrangian from
eleven-dimensional supergravity solution found in \cite{Page}.  
The RG flow along the squashing deformation
trajectory interpolates between ${\cal N}=8$ $SO(8)$-invariant
conformal fixed point at the
IR and ${\cal N}=1$(or ${\cal N}=0$ for different orientation) 
$SO(5) \times SO(3)$-invariant
conformal fixed point
at the UV. 

Then it is natural to ask what happens when we perform 
${\bf Z}_k$-quotient \cite{ABJM} along the above whole RG flow \cite{AR}? 
At the IR fixed point, since the ``round'' ${\bf S}^7$ metric 
is represented by 
a twisted ${\bf S}^1$-bundle over ``round'' ${\bf CP}^3$ characterized by
Fubini-Study Einstein metric \cite{PW,CPW,AI02-1}, one can easily perform 
the ${\bf Z}_k$-quotient. What about at the UV fixed point?
Is there any ``squashed'' ${\bf CP}^3$?
Nilsson and Pope \cite{NP} have realized that 
the ``squashed'' ${\bf CP}^3$ metric can be obtained by taking 
the Hopf fibration of ``round'' ${\bf S}^7$ with its squashed metric from
the observation by Ziller \cite{Ziller}.
However, the explicit form for  the ``squashed'' ${\bf CP}^3$
metric is not presented in \cite{NP} as far as I know and it is not clear how 
one parameter family of ``squashed'' ${\bf CP}^3$ metric, contrary to 
``squashed'' ${\bf S}^7$, arises in order to check the correct
behavior of Ricci tensor in seven-dimensions.

Luckily,
in \cite{AF}, by viewing the ${\bf CP}^3$ as an ${\bf S}^2$-bundle
over ${\bf S}^4$ with the self-dual $SU(2)$ instanton, they reproduced
the generic Ricci tensor for the ${\bf CP}^3$ \cite{NP} and the standard
Fubini-Study Einstein metric(which is Kahler) arises when $\lambda^2=1$ where 
$\lambda$ is a squashing parameter while the second ``squashed'' Einstein
metric(which is nearly Kahler) arises when $\lambda^2=\frac{1}{2}$:
one parameter family of ``squashed'' ${\bf CP}^3$. 

In this paper, we consider the general, one parameter-family, 
metric for ${\bf CP}^3$ 
described in \cite{AF}, 
study its seven-dimensional uplift metric on an ${\bf S}^1$-bundle over this
${\bf CP}^3$, and construct the full eleven-dimensional metric with 
appropriate warp factors describing both breathing mode and squashing
mode.
By analyzing   the scalar potential, 
the holographic supersymmetric(or nonsupersymmetric) 
RG flow 
from ${\cal N}=1$(or ${\cal N}=0$) 
$SO(5) \times U(1)$-invariant UV fixed
point to ${\cal N}=6$(or ${\cal N}=0$ for the opposite orientation) 
$SU(4)_R \times U(1)$-invariant IR fixed point is described
\footnote{Recently, the ${\cal N}=1$ superconformal
Chern-Simons matter theory with $SO(5)\times U(1)$ global symmetry
is constructed in \cite{OP} and they conjecture that 
this is dual to the M-theory on $AdS_4 \times
\widetilde{{\bf S}}^7/{\bf Z}_k$.  }. 

In section 2, we describe the round-quotient(${\bf S}^7/{\bf Z}_k$) and 
squashed-quotient($\widetilde{\bf S}^7/{\bf Z}_k$)  
seven spheres compactification vacua in
eleven-dimensional
supergravity. 
In section 3,
the squashing deformation of  each vacua
is described by an irrelevant operator at the ${\cal N}=6$(or ${\cal N}=0$) 
conformal
fixed point and a relevant operator at the ${\cal N}=1$(or ${\cal N}=0$)
conformal fixed points. The RG flow is described in $AdS_4$
supergravity by a static domain wall interpolating between 
these two vacua. We identify the corresponding operator 
in the boundary conformal field theory.
In appendix A, we present the details for the computations of Ricci
tensor and corresponding field equations.

\section{Round-quotient and squashed-quotient seven spheres }

Let us consider an eleven-dimensional supergravity on $AdS_4 \times
{\bf X}^7$ where ${\bf X}^7$ is a seven-dimensional compact 
Einstein manifold. When the fermion field is set equal to zero, 
the bosonic field equations are given by eleven-dimensional
Einstein equation and Maxwell equation, as usual. 
In order to solve eleven-dimensional Einstein equation for the
Freund-Rubin \cite{FR} form for the gauge field strength, the Ricci
tensor has nonzero components for the indices of four-dimensional
spacetime and the indices of seven-dimensional internal space 
\cite{Page}. For example, see the equations (\ref{result1}).   
On the other hand, one also obtains the Ricci tensor from the 
eleven-dimensional metric (\ref{result}). 

Before we describe the eleven-dimensional metric directly, 
we need to understand the structure of the 
${\bf CP}^3$ internal space metric first. Since we are interested in 
the RG flow connecting two conformal fixed points, it is necessary to 
obtain one parameter family of squashed ${\bf CP}^3$ metric which
allows us to have both round and squashed ${\bf CP}^3$'s we described
in the previous section.  
For the standard Fubini-Study 
metric on the round ${\bf CP}^3$, it contains 
${\bf CP}^2$ \cite{PW,CPW,AI02-1} or ${\bf CP}^1 \times {\bf CP}^1$ 
\cite{CLP,CPW} inside of ${\bf CP}^3$.   
At first sight, it is natural to generalize this standard round 
${\bf CP}^3$ metric to the general one-parameter family ${\bf CP}^3$
by putting the warp factors in front of each orthonormal basis of 
six-dimensional metric. 
However, it does not produce the correct, general 
Ricci tensor components found
in \cite{NP} and there exists only one critical point in the scalar 
potential. 

There is an alternative description for the squashed-seven sphere by 
the fact that  the ${\bf S}^7$ as an ${\bf S}^3$-bundle
over ${\bf S}^4$ with gauge potential for the self-dual $SU(2)$ 
instanton \cite{DNP86,CFLMN}.
Then the squashing is related to the size of the ${\bf S}^3$-fibers
relative to the base ${\bf S}^4$. After an inverse Kaluza-Klein
construction, the seven-dimensional metric consists of
four-dimensional metric plus $SU(2)$ gauge field and 
this metric leads to the metric \cite{ADP,DNP} 
from other approach with quaternionic
projective plane.
Aldazabal and Font \cite{AF}(See also \cite{Tomasiello,KLT}) 
have constructed the family of squashed 
${\bf CP}^3$ metric 
by viewing the ${\bf CP}^3$ as an ${\bf S}^2$-bundle
over ${\bf S}^4$ with the self-dual $SU(2)$ instanton, in the context
of flux compactification on $AdS_4 \times {\bf CP}^3$.
The squashing corresponds to the size of the ${\bf S}^2$-fibers
relative to the base ${\bf S}^4$ which has the line element of
standard Einstein metric on ${\bf S}^4$.
The self-dual $SU(2)$ instanton gauge potential appears 
in the ${\bf S}^2$-bundle.
It turns out that 
the standard
Fubini-Study Einstein metric(which is Kahler) occurs when $\lambda^2=1$ 
while the second squashed Einstein
metric(which is nearly Kahler) occurs when $\lambda^2=\frac{1}{2}$. 
The isometry group corresponding to ${\bf S}^4$ 
in this case is given by $SO(5)$.
See also the relevant paper \cite{GNP} which discusses about the metric
on the higher dimensional case 
${\bf CP}^{2n+1}$ with the quaternionic projective 
space ${\bf HP}^n$.

As noticed in \cite{NP}, the uplift to the seven-dimensional metric
can be done from the expectation that the ${\bf CP}^3$ solution in 
ten-dimensions is originated from the 
seven-sphere ${\bf S}^7$ solution in 
eleven-dimensional supergravity.
By writing the seven-dimensional metric 
on an ${\bf S}^1$-fibration over squashed ${\bf
CP}^3$, one obtains  
the squashed seven-sphere $\widetilde{{\bf S}}^7$ metric.
The gauge potential in the ${\bf S}^1$-fibration  is related to 
the RR 2-form flux \cite{AF}. 
The standard round seven-sphere metric arises when $\lambda^2=1$ 
while the squashed Einstein
metric arises when $\lambda^2=\frac{1}{5}$.
Note that this value is different from the one $\lambda^2=\frac{1}{2}$ 
in squashed 
${\bf CP}^3$.
One can explicitly check that 
the squashed-seven sphere  
$\widetilde{{\bf S}}^7$ metric described as an ${\bf S}^3$-bundle
over ${\bf S}^4$ with gauge potential as we mentioned before
is equivalent to the one on an ${\bf S}^1$-fibration over squashed ${\bf
CP}^3$. This observation is also found in section 4 of \cite{AF}.

Now a general metric that is locally the direct sum of an arbitrary
four-dimensional 
spacetime metric and a squashed seven-sphere metric(obtained before) 
with appropriate
warp factor for the breathing mode \cite{LS}, interpolating between 
round-quotient and squashed-quotient seven spheres,  
may be written as 
\bea
\frac{\overline{ds}^2}{R^2} & = &   
e^{- 7u(x)} g_{\alpha \beta}(x) d x^\alpha d x^\beta
 +      \frac{1}{4} e^{2u(x)+3v(x)}  \left( d \theta^2 + 
\frac{1}{4} \sin^2 \theta 
( \sigma_1^2 + \sigma_2^2 +\sigma_3^2 )   \right) 
 \nonu \\
& + &    \frac{1}{4} e^{2u(x) -4v(x)} \left( d\mu -\sin \phi A^1 + 
\cos \phi  A^2 
\right)^2 \nonu \\
& + &   \frac{1}{4} e^{2u(x)-4v(x)} 
 \sin^2 \mu \left( d \phi - \cot \mu \cos \phi A^1- \cot \mu \sin \phi
 A^2 + A^3 \right)^2 \nonu \\ 
&+ &   \frac{1}{4} e^{2u(x)-4v(x)} \left( \frac{d \psi}{k} -  
\sin \mu \cos \phi A^1 -
   \sin \mu \sin \phi A^2- \cos \mu A^3 
   -\cos \mu d \phi \right)^2, 
\label{11dmetric}
\eea
where the three real one-forms satisfy the $SU(2)$ algebra
$
d \sigma_i = -{1 \over 2} \epsilon_{ijk} \sigma_j \wedge \sigma_k$ and
the self-dual $SU(2)$ instanton gauge potential appearing
in the ${\bf S}^2$-bundle is as follows:
\bea
A^i \equiv \cos^2(\frac{\theta}{2}) \; \sigma_i, \qquad i = 1, 2, 3.
\nonu
\eea
The ${\bf S}^4$ that can be parametrized as $x^5(=\theta),x^6,x^7$ and $x^8$ 
directions has the line element of
standard Einstein metric given in the second term of the first line of 
(\ref{11dmetric}) \footnote{
 The variables we use in this paper can be compared with
  those in \cite{AF} as follows: $\mu \leftrightarrow \theta, 
\phi \leftrightarrow \varphi, \psi \leftrightarrow \tau, 
\theta \leftrightarrow \psi, 
\sigma_i \leftrightarrow \Sigma^i$ and $A^i \leftrightarrow {\cal
  A}^i$. 
By using the relations 
$\Sigma_1 = \sin \phi d \mu + \sin \mu \cos \phi d \psi, \Sigma_2 =
 -\cos \phi d \mu + \sin \mu \sin \phi d \psi$ and 
$ \Sigma_3 = - d \phi + \cos \mu d \psi$ with $k=1$ \cite{AF}, 
one can reexpress the
three-dimensional metric parametrized by $(\mu, \psi, \phi)$ 
as follows: $\lambda^2 
\left( \Sigma_i - A^i \right)^2$. Then it is easy to see that the
whole seven-dimensional metric by adding the other four-dimensional
metric parametrized by $(\theta, x^6, x^7, x^8)$ to this
three-dimensional one gives rise to the usual squashed seven-sphere
metric. 
According to the observation of 
\cite{DNP86}, the squashed-seven sphere  
metric described above as an ${\bf S}^3$-bundle
over ${\bf S}^4$ is equivalent to the standard one found by 
\cite{ADP,Page,AR} 
via the following identifications $\Sigma_i  \leftrightarrow 
\sigma_i$ and $i
\sigma_1 + j \sigma_2 + k \sigma_3 \leftrightarrow 
V( i \omega_1 + j \omega_2 + k
\omega_3) 
V^{-1}$ with quaternion $V$ of unit modulus which can be parametrized
by three Euler angles. Here $i,j,k$ are imaginary quaternions. }. 
The isometry corresponding to ${\bf S}^4$ is given by $SO(5)$.
The ${\bf S}^2$-bundle given in the second 
and the third lines
specifies  the two
coordinates $(\mu, \phi)$ that play the role of $(x^9, x^{10})$ respectively. 
When $v(x)=0$, the isometry corresponding to 
${\bf CP}^3$ becomes $SU(4)$ which is reduced to $SO(5)$ for nonzero $v(x)$.
In the last line, we do the ${\bf Z}_k$-quotient by writing 
\bea
\psi \rightarrow \psi/k, \qquad k > 2.
\label{quo}
\eea
Note that for $k=1$ and $k=2$, the supersymmetry of the IR theory 
is enhanced to ${\cal N}=8$ supersymmetry.
The parameter $R$ measures the overall radius of curvature and the scalar fields 
$u(x)$ and 
$v(x)$ parametrize ``size'' where 
\bea 
\mbox{Vol}({\bf X}^7) = {1 \over 3k} \pi^4 e^{7u} R^7
\label{vol}
\eea 
and ``squashing'' deformation of ${\bf
  X}^7$ over four-dimensional spacetime \footnote{The
  $x$-dependence on $u(x)$ and $v(x)$ only refers to the
  four-dimensional spacetime $x^{\alpha}$ where $\alpha =1, \cdots, 4$.}. 
The squashing is parametrized by 
\bea
\lambda^2(x) \equiv e^{- 7v(x)}.
\label{squ}
\eea

Spontaneous compactification of M-theory to $AdS_4 \times {\bf S}^7$
before ${\bf Z}_k$-quotient is 
obtained from near-horizon geometry of $N$ coincident M2-branes. 
Through the seven-sphere before the ${\bf Z}_k$-quotient,  
the M2-branes thread 
nonvanishing flux of four-form field strength of the Freund-Rubin 
form \cite{FR}:
\bea
\overline{F}_{\alpha \beta \gamma \delta}
= Q' e^{-7 u(x) } \overline{\epsilon}_{\alpha \beta \gamma \delta}
= Q' e^{-21 u(x)} \epsilon_{\alpha \beta \gamma \delta}. 
\label{fieldstrength}
\eea
The Page charge \cite{Page} defined by
\bea
Q' \equiv \pi^{-4} 
\int_{X^7} ({}^*\overline{F} + \overline{C} \wedge \overline{F})
\nonu
\eea 
in the convention of \cite{DNP86}
is related to the total number of
M2-branes, $N'$, as \cite{ABJM} 
\bea
Q' = 96 \pi^2 N' \ell_{\rm p}^6.
\nonu
\eea
Note that the  $\int_{X^7} \overline{C} \wedge \overline{F}$ term above 
vanishes in the Freund-Rubin compactification.
The final number of flux quanta on the ${\bf S}^7/{\bf Z}_k$ after
the ${\bf Z}_k$-quotient (\ref{quo}) is given by \cite{ABJM}
\bea
N = \frac{N'}{k},
\nonu
\eea
because the volume with ${\bf Z}_k$-quotient is smaller by a factor of
$k$ (\ref{vol}) than the original volume without ${\bf Z}_k$-quotient.  
Then the Page charge can be written as
\bea
Q' =  96 \pi^2 N' \ell_{\rm p}^6 =  96 \pi^2  k N \ell_{\rm p}^6 
\equiv k Q, \qquad \mbox{where} \qquad Q = 96 \pi^2 N \ell_{\rm p}^6.
\label{q}
\eea

The four-dimensional field equations from the appendix A can be
obtained from the effective Lagrangian
\bea
{\cal L}= \sqrt{-g} \left[ R - \frac{63}{2} ( \pa u )^2 -
21 ( \pa v )^2 -V(u, v) \right],
\label{Lag}
\eea
where the supergravity scalar potential is given by
\bea
V(u, v)= e^{-9u(x)} \left[ - 6 e^{4v(x)} -48 e^{-3v(x)}+12 e^{-10v(x)} 
+  2 k^2 Q^2 e^{-12u(x)} \right],
\label{potential}
\eea
which depends on $u(x), v(x), k$ and $Q$ with (\ref{q}). 
The $AdS_4$-invariant ground state solutions  
correspond to setting $u=\mbox{const}, v=\mbox{const}$ and the
four-dimensional spacetime curvature is maximally symmetric and 
$R^{\alpha}_{\beta}=\Lambda \delta^{\alpha}_{\beta}$.  

The two vacua with explicit $k$-dependence can be summarized by 
\bea
{\bf S}^7/{\bf Z}_k \,\, \quad : \quad u = u_1 &=& {1 \over 12} \ln
(3^{-2} k^2 Q^2 ),
\qquad v = v_1 = 0, \qquad \qquad(\lambda^2 = 1),
\nonu \\
\Lambda_1 &=& -12 \left\vert { k Q \over 3} \right\vert^{-3/2},
\nonu
\eea
and 
\bea
\widetilde{\bf S}^7/{\bf Z}_k \quad : \quad 
u = u_2 &=& {1 \over 12} \ln (3^{-4} 5^{10/7} k^2 Q^2 ), \qquad
v = v_2 = {1 \over 7} \ln 5, \qquad \left(\lambda^2 = {1 \over 5} \right), 
\nonu \\
\Lambda_2 &=& - 12 \cdot 3^{7/2} 5^{-5/2}
\left\vert {k Q \over 3} \right\vert^{-3/2}.
\nonu
\eea
The two supergravity solutions are classically stable under the
changes of the size and squashing parameter (\ref{squ}) of seven-sphere by
following the analysis found in \cite{Page}.
Note that $u_1$ and $u_2$ depend on $k$ explicitly while 
$v_1$ and $v_2$ do not depend on it.
Obviously, for $k=1$, the theory becomes the original theory without
${\bf Z}_k$-quotient. 
The ${\bf S}^7/{\bf Z}_k$ is a saddle point,  corresponds to a
minimum along the $v$-direction and is invariant under the $SU(4)_R
\times U(1)$ isometry group while $\widetilde{\bf S}^7/{\bf Z}_k$
is a maximum and is invariant under $SO(5) \times U(1)$ subgroup.
Since the ${\bf 8}_s$ of massless gravitini breaks into 
\bea
{\bf 8}_s
\rightarrow {\bf 6}_0 \oplus {\bf 1}_2 \oplus {\bf 1}_{-2} 
\nonu
\eea
under the 
$SU(4)_R \times U(1)$ for the left-handed orientation of the
seven-sphere ${\bf S}^7_L/{\bf Z}_k$,
the near-horizon geometry preserves ${\cal N}=6$ supersymmetry 
\cite{NP,DLP}  because there are six ${\bf Z}_k$-quotient
invariant states.
On the other hand,  the ${\bf 8}_s$ of massless gravitini 
breaks into 
\bea
{\bf 8}_s
\rightarrow {\bf 4}_{-1} \oplus \overline{{\bf 4}}_{1}
\nonu
\eea 
under the 
$SU(4)_R \times U(1)$ for the right-handed orientation of the
seven-sphere  ${\bf S}^7_R/{\bf Z}_k$,
the near-horizon geometry preserves no supersymmetry(${\cal N}=0$) 
at all \cite{NP,DLP} because there is no ${\bf Z}_k$-quotient
invariant state.
Since the left-handed squashed seven-sphere $\widetilde{\bf S}^7_L$ 
gives rise to a theory 
with ${\cal N}=1$ supersymmetry \cite{DNP86}(space invader
scenario-level crossing phenomena among massless and massive
Kaluza-Klein states) and the massless 
gravitino is a singlet under the $U(1)$ in the squashed ${\bf CP}^3$ 
compactification, it will also have ${\cal N}=1$ supersymmetry at
$\widetilde{\bf S}^7_L/{\bf Z}_k$.   
For the right-handed orientation of squashed
seven-sphere $\widetilde{\bf S}^7_R/{\bf Z}_k$,
the near-horizon geometry preserves no 
supersymmetry(${\cal N}=0$) at all.

As stressed in the introduction,  by starting from 
the general, one parameter-family, 
metric for ${\bf CP}^3$ 
described in \cite{AF}, 
and studying its seven-dimensional uplift metric on an ${\bf
S}^1$-bundle 
over this
${\bf CP}^3$, the construction of the full eleven-dimensional metric 
(\ref{11dmetric}) which contains the ``squashed'' ${\bf CP}^3$ with 
appropriate warp factors together is new.
For example, the observation for the nonzero RR 2-form in the 
gauge potential, in the appendix A, 
is not so obvious without knowing the full information
of eleven-dimensional metric. 

\section{(Super)conformal field theories in three dimensions }

Using the results of previous section on the Kaluza-Klein spectrum
under squashing deformations, 
an operator giving rising to a RG flow
associated with the symmetry breaking $SU(4)_R \times U(1) \rightarrow
SO(5)
\times U(1)$ will be identified and 
it turns out the operator is relevant at the 
$\widetilde{\bf S}^7/{\bf Z}_k$ fixed point and irrelevant at the
${\bf S}^7/{\bf Z}_k$ fixed point.

$\bullet$ $SU(4)_R \times U(1)$-invariant conformal fixed point

Let us consider the harmonic 
fluctuations of spacetime metric and $u(x)$ and $v(x)$ scalar fields
around $AdS_4 \times {\bf S}^7/{\bf Z}_k$.
Following \cite{Page}, it turns out more convenient 
to rewrite (\ref{Lag}) in terms of the un-rescaled, 
M-theory metric $\overline{g_{\alpha \beta}} = 
e^{-7u} g_{\alpha \beta}$  in (\ref{11dmetric}) \cite{AR}:
\bea
{\cal L} 
 =   \sqrt{-\overline{g}} e^{7u(x)} \left[ \, \overline{R} -
2\overline{\La_1} - 
105 ( \pa u )^2 - 21 ( \pa v )^2 -2 \overline{V_1}(u, v) \, \right],   
\nonu
\eea 
where the scalar potential is 
\bea
\overline{V_1}(u, v)= 
-\overline{\La_1} \left[ 1-\frac{1}{4} e^{-2(u(x)-u_1)} ( e^{4v(x)}+8e^{-3v(x)}-
2e^{-10v(x)} ) +\frac{3}{4} e^{-14(u(x)-u_1)} \right],
\nonu
\eea
in which the un-rescaled cosmological constant $\overline{\La_1}
= e^{7u_1} \La_1 =  \frac{1}{2} e^{7u_1} V(u_1,v_1)$ is given by 
\bea
\overline{\La_1} \equiv - 12 m_1^2 {1 \over \ell_p^2}= 
-12 \left( \frac{|k Q|}{3} \right)^{-1/3} {1 \over \ell_p^2}
\qquad {\rm where} \qquad m_1 = {1 \over \overline{r_{\rm IR}}}.
\label{lambda}
\eea
Here
$\overline{r_{\rm IR}}$ is related to $N$, $k$ 
and Planck scale $\ell_{\rm p}$ as 
$\overline{r_{\rm IR}} = \ell_{\rm p} {1 \over 2} ( 32 \pi^2 k N)^{1/6}$.
By rescaling the scalar fields as 
\bea
\sqrt{210} u \equiv \overline{u}, \qquad
\sqrt{42} v \equiv \overline{v},
\nonu
\eea 
one obtains that the
fluctuation spectrum for $\overline{v}$-field 
around the ${\bf S}^7/{\bf Z}_k$ 
takes a positive value:
\bea
M_{\overline{v} \overline{v}}^2 ({\bf S}^7/{\bf Z}_k) 
= \left[ \frac{\pa^2} {\pa \overline{v}^2} 2 \overline{V_1}
\right]_{\overline{u}=\overline{u_1}, \overline{v}=\overline{v_1}} = 
-
\frac{4}{3} \overline{\La_1} \ell_p^2=  16 m_1^2.
\label{massmass}
\eea

Recall that before the ${\bf Z}_k$-quotient, 
the $\overline{v}$-field represents squashing of ${\bf
  S}^7$ and hence, under $SO(8)$
isometry group, ought to correspond to ${\bf 300}$ that is the
Young tableaux $\tiny\yng(2,2)$ of $SO(8)$, the lowest mode of the 
transverse, traceless symmetric tensor representation where the  $SO(8)$
Dynkin label is given by $( {\bf 0, 2, 0, 0})$.
The spectrum of supergravity fields is simply the projection of 
the original spectrum on $AdS_4 \times {\bf S}^7$ onto the ${\bf
  Z}_k$-invariant states \cite{NP,Halyo}.
The branching rule of an $SO(8)$ Dynkin label $( {\bf 0, 2, 0, 0}) $ in
terms of ${\bf Z}_k$-invariant 
$SU(4)_R \times U(1)$ Dynkin labels is given
by \cite{NP,Slansky,Petera}  
\bea
({\bf 0, 2, 0, 0})   \rightarrow  ({\bf 0, 0, 0 })\oplus ({\bf
  1, 0, 1} )
\oplus ({\bf 0, 2, 0 }) \oplus ({\bf 2, 0, 2 }) \oplus \cdots, 
\nonu 
\eea
or $SO(8)$ representation ${\bf 300}$ decomposes into 
\bea
{\bf 300}(\tiny\yng(2,2))  \rightarrow   
{\bf 1}(\tiny\yng(1,1,1,1)) \oplus {\bf 15}(\tiny\yng(2,1,1)) 
\oplus {\bf 20'}(\tiny\yng(2,2)) \oplus
{\bf 84}(\tiny\yng(4,2,2)) \oplus \cdots,
\label{su4}
\eea
under the ${\bf Z}_k$-invariant $SU(4)_R \times U(1)$. 
Here, the expressions for $\cdots$ have nonzero $U(1)$ charges and
they are projected out by ${\bf Z}_k$-quotient.
The singlet ${\bf 1}$ corresponds to an overall scaling of the metric, 
$u(x)$. 
Since the ${\bf 20'}$(that is the
Young tableaux $\tiny\yng(2,2)$ of $SU(4)_R$) is represented by a 
traceless symmetric matrix, 
the squashing with ${\bf Z}_k$-quotient
corresponds to nonzero expectation value for the ${\bf 20'}$.
From the mass formula for ${\bf 300}$ \cite{BCERS,AR} from the
eigenvalues of the Lichnerowicz operator, one gets 
\bea 
M_{\bf 20'}^2 = 16 m_1^2,
\nonu
\eea
which is equal to (\ref{massmass}) and note that $m_1^2$ contains 
$k$-dependence from (\ref{lambda}).

One concludes that, in three-dimensional conformal field 
theory with ${\cal N} = 6$ supersymmetry, the $SO(5) \times U(1)$ symmetric 
left-handed squashing should be an irrelevant perturbation 
of conformal dimension 
$\Delta = 4$. Note that this gives a nonsupersymmetric  theory for the 
right-handed squashing seven-sphere orientation ${\bf S}_R^7/{\bf Z}_k$.

$\bullet$ $SO(5) \times U(1)$-invariant conformal fixed point

Due to the skew-whipping, the theory will be either 
left squashed quotient $\widetilde{\bf S}^7_L/{\bf Z}_k$ with ${\cal
  N}=1$ supersymmetry or 
right squashed quotient $\widetilde{\bf S}^7_R/{\bf Z}_k$ with ${\cal
  N}=0$ supersymmetry.
Note that the ${\bf Z}_k$-quotient acts on the $U(1)$ subgroup 
of $SU(2)$ which is present when there is no ${\bf Z}_k$-quotient. 
The isometry of the squashed seven-sphere is $SO(5) \times
SU(2)$ and this is broken to $SO(5) \times U(1)$ 
by the ${\bf Z}_k$ quotient.
In terms of the un-rescaled M-theory metric, the Lagrangian 
(\ref{Lag}) may be rewritten as
\bea
{\cal L}  = 
\sqrt{-\overline{g}} e^{7u(x)} \left[ \, \overline{R} -
2\overline{\La_2}  - 
105 ( \pa u )^2 -
21 ( \pa v )^2 -2 \overline{V_2}(u, v) \, \right],   
\nonu
\eea 
where the scalar potential is given by
\bea
\overline{V_2}(u, v)  = 
 -\overline{\La_2} \left[ 1-\frac{1}{36} e^{-2(u(x)-u_2)} 
\left( 25 e^{4(v(x)-v_2)}+ 40 e^{-3(v(x)-v_2)}- 2 e^{-10(v(x)-v_2)} \right)  
+\frac{3}{4} e^{-14(u(x)-u_2)} \right],
\nonu
\eea
and the un-rescaled cosmological constant $\overline{\La_2} =
e^{7 u_2} \La_2 =\frac{1}{2} e^{7u_2} V(u_2,v_2)$ is given by
\bea
\overline{\La_2}  &\equiv& - 12 m_2^2 {1 \over \ell_{\rm p}^2}
= - 12 \cdot 3^{7/3} 5^{-5/3}  \left( {\vert k Q \vert  \over 3} \right)^{-1/3} 
{1 \over \ell_{\rm p}^2}, \quad {\rm where}
\quad m_2 = {1 \over \overline{ r_{\rm UV}}   }.
\nonu
\eea
Once again, the mass spectrum of the $\overline{v}(x)$ field is calculated 
straightforwardly:
\bea
M^2_{\overline{v} \overline{v}}[\widetilde{\bf S}^7/{\bf Z}_k] 
\equiv \left[ \frac{\pa^2}
{\pa \overline{v}^2} 2 \overline{V_2} \right]_{\overline{u}=
\overline{u_2}, \overline{v}=\overline{v_2}} \,\, = \,\,  
\frac{20}{27} \overline{\La_2} \ell^2_{\rm p} = -\frac{80}{9} m_2^2.
\label{mass2}
\eea

The branching rule of the ${\bf Z}_k$-invariant 
$SU(4)_R$ representations appearing in the right hand side of (\ref{su4}) in
terms of $SO(5)$ representation is given by  
\bea
{\bf 1} & \rightarrow & {\bf 1}, \nonu \\
{\bf 15} & \rightarrow & {\bf 5} \oplus {\bf 10}, \nonu \\
{\bf 20'} & \rightarrow & {\bf 1} \oplus {\bf 5} \oplus {\bf 14},
\nonu \\ 
{\bf 84} & \rightarrow & {\bf 14} \oplus {\bf 35} \oplus {\bf 35}.
\nonu
\eea
Similar aspect in the branching rule of ${\bf 20'}$ 
for $AdS_5 \times {\bf S}^5$ compactification 
has been found in \cite{GPPZ,DZ} in the context of gauged supergravity.
In particular, the singlet ${\bf 1}$ in the decomposition of 
${\bf 20'}$ corresponds to $\overline{v}(x)$ field we turned on. From the 
mass formula for the $SO(5) \times SU(2)$ 
representation(before ${\bf Z}_k$-quotient)
\cite{DNP86,AR}
from the
eigenvalues of the Lichnerowicz operator, 
one obtains the mass-squared for the singlet ${\bf 1}$ 
as follows:
\bea
M_{\bf 1}^2 =  -\frac{80}{9} m_2^2,
\nonu
\eea
and this coincides with (\ref{mass2}).
The perturbation that corresponds to squashing
around $\widetilde{ \bf S}^7/{\bf Z}_k$ has a scaling dimension either 
$\Delta = 4/3$ or $5/3$ and hence corresponds to a relevant operator.

We gave a nonzero expectation value to a supergravity scalar in the 
${\bf 20'}$ of $SU(4)_R$. Using the AdS/CFT correspondence, one
identifies this perturbation with a composite operator of 
${\cal N}=6$ superconformal Chern-Simons matter theory with a mass
term for the symmetric and traceless product between 
two ${\bf 6}$'s: $\lambda_{AB} \int d^3 x {\cal O}^{AB}$ where 
$\lambda_{AB}$ is in the ${\bf 20'}$ of ${\bf Z}_k$-invariant $SU(4)_R$. 
Note that the tensor product of these leads to
\bea
{\bf 6}(\tiny\yng(1,1)) \times {\bf 6}(\tiny\yng(1,1)) = 
{\bf 1}(\tiny\yng(1,1,1,1)) 
\oplus {\bf 15}(\tiny\yng(2,1,1)) \oplus {\bf 20'}(\tiny\yng(2,2))
\nonu
\eea 
in $SU(4)_R$ 
representation. Then one can construct a ${\bf 6}$(that is the
Young tableaux $\tiny\yng(1,1)$)
representation by
using the Clebsch-Gordan coefficient $\Gamma^A_{IJ}(A=1, \cdots, 6)$ 
\cite{BLS1,OP} 
which transforms
two ${\bf 4}$'s into ${\bf 6}$ of $SU(4)_R$ together with matter field
$C^I$: 
\bea
\Gamma^A_{IJ} C^I C^J
\nonu
\eea
where
$C^I(I=1, \cdots, 4)$ are four
complex scalars(${\bf 4}$ under the $SU(4)_R$) 
transforming as $({\bf N, \overline{N}})$ with gauge
group $U(N) \times U(N)$ in ${\cal N}=6$ superconformal Chern-Simons
gauge theory \cite{ABJM}.  
The perturbation breaking $SU(4)_R \times U(1)$ to 
$SO(5) \times U(1)$ is given by 
\bea
{\cal O}^{AB} = \Tr \Gamma^A_{IJ} C^I
C^J \Gamma^B_{KL} C^K C^L - \frac{1}{6} \delta^{AB}  \Tr \Gamma^C_{IJ} C^I
C^J \Gamma^C_{KL} C^K C^L.
\label{operator}
\eea
The singlet ${\bf 1}$ 
of this operator ${\cal O}^{AB}$ which is ${\bf 20'}$ 
of $SU(4)_R$ corresponds to the
supergravity
field $v(x)$(or $\overline{v}(x)$) and the conformal dimensions are
given by $\Delta_{UV} = \frac{4}{3}$(or $\frac{5}{3}$) and $\Delta_{IR}=4$  
respectively as we computed before.

The massless supermultiplets 
contain $(2-\frac{3}{2})$-supermultiplet in the $SU(4)$ singlet,
$(1-\frac{1}{2})$-supermultiplet in the ${\bf 10} \oplus {\bf 1}$ of
$SU(4)$ \cite{NP}.
The full spectrum of massive supermultiplets is the subset of 
the multiplets on the left squashed seven sphere that are neutral 
when $SU(2)$ breaks to $U(1)$. 
There exist ${\bf 35} \oplus {\bf 30}$ of $SO(5)$ massless scalars
in massive supermultiplets which are the members of massive
Wess-Zumino multiplets.  

Since the scalar potential we are considering is almost the same as
the one in \cite{AR} by replacing $Q$ with $k Q$, 
the non-perturbative analysis for the stability 
given in \cite{AR}, besides the perturbative analysis considered so far,
appears as a static domain wall. 
With the equations of motion for the metric, $u$ and $v$, the
four-dimensional metric ansatz satisfies  the correct boundary
conditions at UV and IR regions. Then the Ricci tensor and the Ricci
scalar can be determined in this background. Using the asymptotics, 
one can check the consistency of the $SO(3)$ symmetric domain-wall
configuration.
One sees the rescaled $u$ field has $Q$ dependence which can be
rescaled by $k Q$ at the UV and IR regions. One obtains the regular
asymptote
of the ratio of the scalar potential at two fixed points with
monotonic rescaling of the curvature of radius. This monotonic 
radial behavior of the static $SO(3)$ domain-wall configuration is the 
holographic representation of the renormalization group flow 
\cite{GPPZ,PS,FGPW}.

$\bullet$ $SU(2) \times SU(2) \times U(1)$-invariant conformal fixed point

So far, we have considered a particular one-parameter RG
flow between
conformal fixed points of M2-brane worldvolume theory. Geometrically,
the flow is induced from varying the position of M2-brane when placed 
near a conical singularity of an eight-dimensional manifold with 
$Spin(7)$ holonomy. In the IR limit, the conical singularity and
hence squashing of M2-brane horizon are washed out completely. 

At the IR fixed point with $SU(4)_R \times U(1)$ 
symmetry, one may flow further into 
another fixed points by turning on a set of relevant operators. 
It includes 
scalar operators of Dynkin label $({\bf n, 0, 0, 0})$ with ${\bf n} \geq 2$
and pseudoscalar operators of Dynkin label $({\bf n, 0, 2, 0})$ with
${\bf n} \geq 0$. 
Among them are 70 scalar fields ${\bf 35}_{\rm v}(\tiny\yng(2)) \oplus
{\bf 35}_{\rm c}(\tiny\yng(1,1,1,1))$ 
of $SO(8)$ in the massless gravity supermultiplet,
parametrizing the coset space $E_{7(7)}/SU(8)$ in ${\cal N}=8$ gauged 
supergravity. 
Decomposing them under the ${\bf Z}_k$-invariant 
$SU(4)_R \times U(1) \subset SO(8)_R$ \cite{NP,DLP},
\bea
{\bf 35}_{\rm v} \oplus {\bf 35}_{\rm c} \rightarrow
{\bf 15} \oplus {\bf 15} \oplus \cdots 
\nonu
\eea 
where we present only the
representations with vanishing $U(1)$ charge. 
Turning on the two relevant operators
${\bf 15}$'s breaks ${\bf Z}_k$-invariant 
$SU(4)_R \times U(1) \rightarrow SU(2) \times SU(2)
\times U(1)$ and they have the following branching rule
\bea
{\bf 15} \oplus {\bf 15} 
\rightarrow ({\bf 1, 1}) \oplus ({\bf 1, 1}) \oplus \cdots
\nonu
\eea
where we write only the representations with vanishing $U(1)$ charge.
Utilizing
on the known result, in \cite{DZ}, they
have studied RG flow to a nonsupersymmetric vacuum 
with $SU(2) \times SU(2) \times U(1)$ global symmetry.  

Moreover, in the context of three-dimensional boundary theory 
\cite{GRVV}, 
the $SU(2) \times SU(2)$ singlet $({\bf 1, 1})$ is found from
the operator ${\bf 15}$(which is the Young tableaux $\tiny\yng(2,1,1)$),
appearing in the branching rule of ${\bf 35}_v$, that corresponds to 
\cite{ABJM} 
\bea
{\cal P}^{I}_{J} \sim \Tr C^I 
C^{\dagger}_{J} + \cdots.
\label{15}
\eea
Here
$C^{\dagger}_J(J=1, \cdots, 4)$ are four
complex scalars(${\bf \overline{4}}$ under the $SU(4)_R$) 
transforming as $({\bf \overline{N}, N})$ with gauge
group $U(N) \times U(N)$ in ${\cal N}=6$ superconformal Chern-Simons
gauge theory \cite{ABJM}.
On the other hand,
the other $SU(2) \times SU(2)$ singlet $({\bf 1, 1})$ is 
found from the operator ${\bf 15}$,
appearing in the branching rule of ${\bf 35}_c$, written as \cite{GRVV}
\bea
{\cal O}^{I}_{J} \sim \Tr C^K C^{\dagger}_{[K} C^I 
C^{\dagger}_{J]} + \cdots. 
\nonu
\eea
This is obtained from the one (\ref{15}) by acting two supersymmetry
transformations which are in the ${\bf 6}$ of $SU(4)_R$. 

\section{
Conclusions and outlook }

We have constructed the full eleven-dimensional metric given by 
(\ref{11dmetric}) and obtained  the scalar potential in 
(\ref{potential}) by using the Freund-Rubin ansatz
(\ref{fieldstrength}) with the help of appendix A.  
The holographic supersymmetric(or nonsupersymmetric) RG flow 
from ${\cal N}=1$(or ${\cal N}=0$) 
$SO(5) \times U(1)$-invariant UV fixed
point to ${\cal N}=6$(or ${\cal N}=0$) 
$SU(4)_R \times U(1)$-invariant IR fixed point was described.
The corresponding operator in three-dimensional 
Chern-Simons matter theories is characterized by (\ref{operator}).

When there are gauge potential components which are proportional to
totally antisymmetric torsion tensor on the seven sphere as well as
the Freund-Rubin components we have discussed so far, 
there is a Lorentz scalar field $w=w(x)$ providing these gauge potential
components \cite{Page}. 
After analyzing the eleven-dimensional Maxwell and Einstein 
equations, the effective four-dimensional scalar potential is obtained.
However, this has no extremum at finite $u(x), w(x)$ if $Q'$ is
positive
but if $Q'$ is negative, there exists a single extremum at nonzero
$u(x)$
and $w(x)=0$.  
One finds that there is no extra critical point except this single
extremum and concludes that there ought to be no nontrivial RG group
flow in the dual three-dimensional conformal field theory. 
Similar phenomena can be found in \cite{AR1}.

For the seven-dimensional metric, we have considered $SU(2)$-bundle over
the base ${\bf S}^4$ in (\ref{11dmetric}). 
Also one replaces the $SU(2)$-bundle with 
$SO(3)$-bundle or the base ${\bf S}^4$ can be replaced by either ${\bf
CP}^2$ or ${\bf CP}^1 \times {\bf CP}^1$. It is interesting to 
find out whether these metrics provide the nontrivial eleven-dimensional 
solutions and whether there exist nontrivial critical points in the effective 
four-dimensional theory.  
When the ${\bf CP}^3$ in \cite{PW,CPW,AI02-1} is generalized to the
present form, a family of squashed ${\bf CP}^3$, 
what happens for the eleven-dimensional solutions in the
gauged supergravity?

So far, it is known that 
there exist only for ${\cal N}=1, 2$ supersymmetric RG flows 
\cite{Ahn08-1,Ahn08} dual to the Chern-Simons gauge theories with mass
deformation. It is an open problem whether 
${\cal N}=3, 4, 5, 6, 8$ supersymmetric \cite{AFHS,MP} 
RG flows exist in the context of 
${\cal N}=8$ gauged supergravity. 
Are there any new general $AdS_4$ vacua? Are there any new critical
points in the context of $SO(5)$ gauged supergravity 
which might be related to 
${\cal N}=5$ Chern-Simons matter theory or 
$SO(6)$ gauged supergravity?
Are there any flows connecting $AdS_4 \times {\bf S}^7/{\bf Z}_k$ to 
$AdS_5 \times T^{1,1}$ which was suggested in \cite{ABJM}?

There are much progress \cite{FHPR}-\cite{BLS} 
on the direction of \cite{ABJM}. It would be
interesting to apply the findings of this paper to them and see
whether there exist any nontrivial aspects. 
For example, one considers the general family of 
squashed ${\bf CP}^3$ metric described
in this paper and it is interesting to see how the squashing parameter 
$\lambda$(or $v(x)$) 
arises in the $AdS_4 \times {\bf CP}^3$ compactification.

\vspace{.7cm}
\centerline{\bf Acknowledgments}

I would like to thank 
K. Woo for mathematica computations and
O. Lunin  and D. Kutasov for discussions.
This work was supported by grant No.
R01-2006-000-10965-0 from the Basic Research Program of the Korea
Science \& Engineering Foundation.  
I acknowledge warm hospitality of 
Particle Theory Group, Enrico Fermi Institute 
at University of Chicago
where this work was initiated.

\appendix

\renewcommand{\thesection}{\large \bf \mbox{Appendix~}\Alph{section}}
\renewcommand{\theequation}{\Alph{section}\mbox{.}\arabic{equation}}

\section{The Ricci tensor and field equations}

The Ricci tensor \cite{DNP86}
can be obtained from connection one-form and curvature two-form in the
following orthonormal basis for the metric (\ref{11dmetric})
\bea
e^1 & = &   e^{- \frac{7}{2} u(x)} \sqrt{g_{11}(x)} dx^1, \nonu \\
e^2 & = &  e^{- \frac{7}{2} u(x)} \sqrt{g_{22}(x)} dx^2, \nonu \\
e^3 & =  &  e^{- \frac{7}{2} u(x)} \sqrt{g_{33}(x)} dx^3, \nonu \\
e^4 & = &   e^{- \frac{7}{2} u(x)} \sqrt{g_{44}(x)} dx^4, \nonu \\
e^5 & = &  \frac{1}{2} e^{u(x)+\frac{3}{2} v(x)}  d \theta, \nonu \\
e^6 & = &  \frac{1}{2} e^{u(x)+\frac{3}{2} v(x)} 
\sin \theta \sigma_1, \nonu \\
e^7 & = &  \frac{1}{2} e^{u(x)+\frac{3}{2} v(x)} \sin \theta 
\sigma_2, \nonu \\
e^8 & = &  \frac{1}{2} e^{u(x)+\frac{3}{2} v(x)} \sin \theta 
\sigma_3, \nonu \\  
e^9 & = & \frac{1}{2} e^{u(x) -2v(x)} 
\left( d\mu -\sin \phi A^1 + \cos \phi  A^2 \right), \nonu \\
e^{10} & = & \frac{1}{2} e^{u(x)-2v(x)} \sin \mu 
\left( d \phi - \cot \mu \cos \phi A^1- \cot \mu \sin \phi A^2 + A^3
\right), 
\nonu \\
e^{11} & = & \frac{1}{2} e^{u(x)-2v(x)} 
\left( \frac{d \psi}{k} -  \sin \mu \cos \phi A^1 - \sin \mu \sin 
\phi A^2- \cos \mu A^3 
   -\cos \mu d \phi \right), 
\label{frame}
\eea
where  the three real one-forms are 
\bea
\sigma_1  & = & \cos x^8 d x^6 + \sin x^8 \sin x^6 d x^7, \nonu \\ 
\sigma_2  & = & -\sin x^8 d x^6 + \cos x^8 \sin x^6 d x^7, \nonu \\
\sigma_3  & = & d x^8 + \cos x^6 d x^7,
\nonu
\eea
and 
the self-dual $SU(2)$ instanton gauge potential is 
$ A^i \equiv \cos^2(\frac{\theta}{2}) 
\; \sigma_i$. Intentionally, we inserted the overall factor
$\frac{1}{2}$ in the seven-dimensional internal space. 
The RR 2-form is given by the gauge potential in $e^{11}$ through 
\bea
&& d(  \lambda \sin \mu \cos \phi A^1 + \lambda 
\sin \mu \sin \phi A^2+ \lambda \cos \mu A^3 
   +\lambda \cos \mu d \phi)= 
 -\lambda \sin \mu \cos \phi (e^1 \wedge e^2+e^3
   \wedge e^4) \nonu \\
&& -\lambda \sin \mu \sin \phi(e^1 \wedge e^3 -e^2 \wedge
   e^4)-\lambda \cos \mu (e^1 \wedge e^4 + e^2 \wedge
   e^3)-\frac{1}{\lambda} e^5 \wedge e^6,
\nonu
\eea
where the nonzero components for $F_{13}, F_{24}, F_{14}$ and $F_{23}$ 
also occur, contrary to the case of \cite{NP} in which there exist
only nonzero components for $F_{12}, F_{34}$ and $F_{56}$.

Starting from the gauge field strength (\ref{fieldstrength}), 
the eleven-dimensional Einstein equation implies the following Ricci
tensor components:
\bea
\overline{R}^{\alpha}_{\beta} = -\frac{4}{3} Q'^2 e^{-14 u(x)} 
\delta^{\alpha}_{\beta}, \qquad 
\overline{R}^{a}_{b} = \frac{2}{3} Q'^2 e^{-14 u(x)} 
\delta^{a}_{b},  \qquad
\overline{R}^{\alpha}_{b} = \overline{R}^{a}_{\beta} = 0.
\label{result1}
\eea
 
Instead of using the differential forms in orthonormal basis 
in order to get the 
Ricci tensor components by hands, we compute them by using a
mathematica in the basis $dx^i$
where $i=1, 2, \cdots, 11$ and then convert them into the orthonormal 
basis (\ref{frame}).  
The metric connection can be obtained from the eleven-dimensional
metric and Riemann tensor can be determined from this metric
connection.
Finally, the Ricci tensor is defined by the contraction between this
Riemann tensor and eleven-dimensional metric.
Furthermore, using the orthonormal basis (\ref{frame}), one obtains
the following Ricci tensor components in the orthonormal basis 
(\ref{frame})  
\bea
\overline{R}^{\alpha}_{\beta}  & = &  e^{7u(x)} \left[ R^\alpha_\beta +
\frac{7}{2} \delta_{\beta}^{\alpha} u^{;\gamma}_{;\gamma}(x)
-\frac{63}{2}
u^{;\alpha}(x) u_{;\beta}(x) -21 v^{;\alpha}(x) v_{;\beta}(x) \right],
\nonu \\
\overline{R}^{5}_{5} & = & 12 e^{-2u(x)-3v(x)} - 6 e^{-2u(x)-10v(x)} -
e^{7u(x)} \left[ u^{;\alpha}_{;\alpha}(x) +
\frac{3}{2} v^{;\alpha}_{;\alpha}(x) \right]
=\overline{R}^{6}_{6} =\overline{R}^{7}_{7}=\overline{R}^{8}_{8},
\nonu \\
\overline{R}^{9}_{9} & = & 2 e^{-2u(x)+4v(x)} + 4 e^{-2u(x)-10v(x)} -
e^{7u(x)} \left[ u^{;\alpha}_{;\alpha}(x) -2 v^{;\alpha}_{;\alpha}(x) \right]
=\overline{R}^{10}_{10} =\overline{R}^{11}_{11}.
\label{result}
\eea
All these expressions are the same as the one in \cite{Page}. Here 
$ u^{;\alpha} = g^{\alpha \beta} u_{;\beta}$ where the semicolon stands
for the covariant
derivative which contains the metric connection, as usual. 
When $u(x)$ and $v(x)$ are constant, then 
\bea
\overline{R}^{5}_{5} & = & 3 - \frac{3}{2} e^{-7v}=3 - 
\frac{3}{2} \lambda^2 =\overline{R}^{6}_{6} =\overline{R}^{7}_{7}=
\overline{R}^{8}_{8},  \nonu \\
\overline{R}^{9}_{9} & = & e^{-7v} +\frac{1}{2} e^{7v} = \lambda^2 +
\frac{1}{2\lambda^2} = \overline{R}^{10}_{10} =\overline{R}^{11}_{11},
\label{constRicci}
\eea 
except the
overall factor $4 e^{-2u-3v}$.
These (\ref{constRicci}) 
are the same as the one in \cite{ADP,DNP}.
The overall factor comes from different normalization in the metric.
If we substitute $e^{2u} = \frac{1}{4} e^{-3v}$ into
(\ref{11dmetric}), 
then this factor becomes one and the normalization for the
seven-dimensional metric here becomes the standard one \cite{ADP,DNP}.
Then the Einstein condition is satisfied by two values of $\lambda^2$.
The round sphere has $\lambda^2=1$ while the squashed sphere has 
$\lambda^2=\frac{1}{5}$ \footnote{Let us recall that the Ricci tensor
of the ${\bf CP}^3$  has the following components:
$\overline{R}^{5}_{5} = 3 - \lambda^2 =\overline{R}^{6}_{6} =\overline{R}^{7}_{7}=
\overline{R}^{8}_{8}$ and $\overline{R}^{9}_{9} = \lambda^2 +
\frac{1}{\lambda^2} = \overline{R}^{10}_{10}$ \cite{NP,AF} when we
consider the six-dimensional internal space only.  The standard
Fubini-Study Einstein metric arises when $\lambda^2=1$ 
while the second squashed Einstein
metric arises when $\lambda^2=\frac{1}{2}$ as we mentioned in section 1.}. 

Now substituting the last two relations of (\ref{result}) into 
the first two relations of (\ref{result1}) leads to 
the field equations for $u(x)$ and $v(x)$
\bea
 u^{;\alpha}_{;\alpha}(x)  & = & 
\frac{6}{7} e^{-9u(x)+4v(x)} +\frac{48}{7} e^{-9u(x)-3v(x)} -
\frac{12}{7} e^{-9u(x)-10v(x)}-
\frac{2}{3} Q'^2 e^{-21u(x)}, \nonu \\
v^{;\alpha}_{;\alpha}(x)  & = & -\frac{4}{7} e^{-9u(x)+4v(x)} +\frac{24}{7} 
e^{-9u(x)-3v(x)} -\frac{20}{7} e^{-9u(x)-10v(x)}.
\label{result2}
\eea
The second equation of (\ref{result2}) implies 
that either $v_1=0$ or $v_2=\frac{1}{7} \ln 5$ in section 2.
Moreover, plugging the first equation of 
(\ref{result2}) into the first equation of (\ref{result}) and equating
this to the right hand side of the first equation of (\ref{result1}) provides 
\bea
R^{\alpha}_{\beta} & = & {63 \over 2} u^{;\alpha}(x) u_{;\beta}(x)
+ 21 v^{;\alpha}(x) v_{;\beta}(x) \nonu \\
&+& \delta^{\alpha}_{\beta} e^{-9u(x)}
\left[ - 3 e^{+4v(x)} - 24 e^{-3v(x)} +6 e^{- 10v(x)} + 
Q'^2 e^{-12u(x)} \right].
\label{result3}
\eea
Then it is easy to see that the field equations (\ref{result2}) and 
(\ref{result3}) are equivalent to the Euler-Lagrange equations for the 
effective Lagrangian (\ref{Lag}).
When $u(x)$ and $v(x)$ are constant, then $R^{\alpha}_{\beta}  
=\Lambda \delta^{\alpha}_{\beta}=\frac{1}{2} V  \delta^{\alpha}_{\beta}$.


\begin{thebibliography}{99}

\bibitem{ABJM}
  O.~Aharony, O.~Bergman, D.~L.~Jafferis and J.~Maldacena,
  arXiv:0806.1218 [hep-th].

\bibitem{MP}
  D.~R.~Morrison and M.~R.~Plesser,
  Adv.\ Theor.\ Math.\ Phys.\  {\bf 3}, 1 (1999).

\bibitem{Halyo}
  E.~Halyo,
  Mod.\ Phys.\ Lett.\  A {\bf 15}, 397 (2000).

\bibitem{FKPZ}
  S.~Ferrara, A.~Kehagias, H.~Partouche and A.~Zaffaroni,
  Phys.\ Lett.\  B {\bf 431}, 42 (1998).

\bibitem{Gomis}
  J.~Gomis,
  Phys.\ Lett.\  B {\bf 435}, 299 (1998).

\bibitem{AOT}
  C.~Ahn, K.~Oh and R.~Tatar,
  JHEP {\bf 9811}, 024 (1998).

\bibitem{NP}
  B.~E.~W.~Nilsson and C.~N.~Pope,
  Class.\ Quant.\ Grav.\  {\bf 1}, 499 (1984).

\bibitem{BL}
  J.~Bagger and N.~Lambert,
  JHEP {\bf 0802}, 105 (2008).

\bibitem{BL1}
  J.~Bagger and N.~Lambert,
  Phys.\ Rev.\  D {\bf 77}, 065008 (2008).

\bibitem{BL2}
  J.~Bagger and N.~Lambert,
  Phys.\ Rev.\  D {\bf 75}, 045020 (2007).

\bibitem{AR}
  C.~Ahn and S.~J.~Rey,
  Nucl.\ Phys.\  B {\bf 565}, 210 (2000).

\bibitem{Jensen} G. Jensen, J. Diff. Geom. {\bf 8} (1973) 599.

\bibitem{BK}
J.~P.~Bourguignon and H. Karcher, Ann. Sci. Normale Sup. {\bf 11}
(1978) 71.

\bibitem{ADP}
  M.~A.~Awada, M.~J.~Duff and C.~N.~Pope,
  Phys.\ Rev.\ Lett.\  {\bf 50}, 294 (1983).

\bibitem{DNP}
  M.~J.~Duff, B.~E.~W.~Nilsson and C.~N.~Pope,
  Phys.\ Rev.\ Lett.\  {\bf 50}, 2043 (1983); {\bf 51}, 846 (1983) (errata).

\bibitem{Maldacena}
  J.~M.~Maldacena,
  Adv.\ Theor.\ Math.\ Phys.\  {\bf 2}, 231 (1998)
  [Int.\ J.\ Theor.\ Phys.\  {\bf 38}, 1113 (1999)].

\bibitem{Witten}
  E.~Witten,
  Adv.\ Theor.\ Math.\ Phys.\  {\bf 2}, 253 (1998).

\bibitem{GKP}
  S.~S.~Gubser, I.~R.~Klebanov and A.~M.~Polyakov,
  Phys.\ Lett.\  B {\bf 428}, 105 (1998).

\bibitem{Page}
  D.~N.~Page,
  Phys.\ Rev.\  D {\bf 28}, 2976 (1983).

\bibitem{PW}
  C.~N.~Pope and N.~P.~Warner,
  Phys.\ Lett.\  B {\bf 150}, 352 (1985).

\bibitem{CPW}
  R.~Corrado, K.~Pilch and N.~P.~Warner,
  Nucl.\ Phys.\  B {\bf 629}, 74 (2002).

\bibitem{AI02-1}
  C.~Ahn and T.~Itoh,
  Nucl.\ Phys.\  B {\bf 646}, 257 (2002).

\bibitem{Ziller}
W. Ziller, 
Math. Ann. {\bf 259}, 351 (1982).

\bibitem{AF}
  G.~Aldazabal and A.~Font,
  JHEP {\bf 0802}, 086 (2008).

\bibitem{OP}
  H.~Ooguri and C.~S.~Park,
  arXiv:0808.0500 [hep-th].

\bibitem{FR}
  P.~G.~O.~Freund and M.~A.~Rubin,
  Phys.\ Lett.\  B {\bf 97}, 233 (1980).

\bibitem{CLP}
  M.~Cvetic, H.~Lu and C.~N.~Pope,
  Nucl.\ Phys.\  B {\bf 597}, 172 (2001).

\bibitem{DNP86}
  M.~J.~Duff, B.~E.~W.~Nilsson and C.~N.~Pope,
  Phys.\ Rept.\  {\bf 130}, 1 (1986).

\bibitem{CFLMN}
  V.~L.~Campos, G.~Ferretti, H.~Larsson, D.~Martelli and B.~E.~W.~Nilsson,
  JHEP {\bf 0006}, 023 (2000).

\bibitem{Tomasiello}
  A.~Tomasiello,
  Phys.\ Rev.\  D {\bf 78}, 046007 (2008).

\bibitem{KLT}
  P.~Koerber, D.~Lust and D.~Tsimpis,
  JHEP {\bf 0807}, 017 (2008).

\bibitem{GNP}
  G.~W.~Gibbons, D.~N.~Page and C.~N.~Pope,
  Commun.\ Math.\ Phys.\  {\bf 127}, 529 (1990).

\bibitem{LS}
  J.~T.~Liu and H.~Sati,
  Nucl.\ Phys.\  B {\bf 605}, 116 (2001).

\bibitem{DLP}
  M.~J.~Duff, H.~Lu and C.~N.~Pope,
  Phys.\ Lett.\  B {\bf 409}, 136 (1997).


\bibitem{Slansky}
  R.~Slansky,
  Phys.\ Rept.\  {\bf 79}, 1 (1981).

\bibitem{Petera} 
J. Patera and D. Sankoff, {\sl Tables of Branching
Rules for 
Representations of Simple Lie Algebras}(L'Universit\'{e} de Montr\'{e}al, 
Montr\'{e}al, 1973); 
W. MacKay and J. Petera, {\sl Tables of Dimensions, Indices and 
Branching Rules for representations of Simple Algebras}(Dekker, New
York, 1981).

\bibitem{BCERS}
  B.~Biran, A.~Casher, F.~Englert, M.~Rooman and P.~Spindel,
  Phys.\ Lett.\  B {\bf 134}, 179 (1984).

\bibitem{GPPZ}
  L.~Girardello, M.~Petrini, M.~Porrati and A.~Zaffaroni,
  JHEP {\bf 9812}, 022 (1998).

\bibitem{DZ}
  J.~Distler and F.~Zamora,
  Adv.\ Theor.\ Math.\ Phys.\  {\bf 2}, 1405 (1999).

\bibitem{BLS1}
  M.~A.~Bandres, A.~E.~Lipstein and J.~H.~Schwarz,
  JHEP {\bf 0809}, 027 (2008).

\bibitem{PS}
  M.~Porrati and A.~Starinets,
  Phys.\ Lett.\  B {\bf 454}, 77 (1999).

\bibitem{FGPW}
  D.~Z.~Freedman, S.~S.~Gubser, K.~Pilch and N.~P.~Warner,
  Adv.\ Theor.\ Math.\ Phys.\  {\bf 3}, 363 (1999).

\bibitem{GRVV}
  J.~Gomis, D.~Rodriguez-Gomez, M.~Van Raamsdonk and H.~Verlinde,
  arXiv:0807.1074 [hep-th].

\bibitem{AR1}
  C.~Ahn and S.~J.~Rey,
  Nucl.\ Phys.\  B {\bf 572}, 188 (2000).

\bibitem{Ahn08-1}
  C.~Ahn,
  JHEP {\bf 0807}, 101 (2008).

\bibitem{Ahn08}
  C.~Ahn,
  JHEP {\bf 0808}, 083 (2008).

\bibitem{AFHS}
  B.~S.~Acharya, J.~M.~Figueroa-O'Farrill, C.~M.~Hull and B.~J.~Spence,
  Adv.\ Theor.\ Math.\ Phys.\  {\bf 2}, 1249 (1999).



\bibitem{FHPR}
  S.~Franco, A.~Hanany, J.~Park and D.~Rodriguez-Gomez,
  arXiv:0809.3237 [hep-th].

\bibitem{KP}
  J.~Kluson and K.~L.~Panigrahi,
  arXiv:0809.3355 [hep-th].

\bibitem{Chen:2008vc}
  C.~M.~Chen, J.~H.~Tsai and W.~Y.~Wen,
  arXiv:0809.3269 [hep-th].

\bibitem{LLP}
  K.~Lee, S.~Lee and J.~H.~Park,
  arXiv:0809.2924 [hep-th].

\bibitem{Chen:2008bp}
  B.~Chen and J.~B.~Wu,
  arXiv:0809.2863 [hep-th].

\bibitem{Drukker:2008zx}
  N.~Drukker, J.~Plefka and D.~Young,
  arXiv:0809.2787 [hep-th].

\bibitem{HTT}
  K.~Hashimoto, T.~S.~Tai and S.~Terashima,
  arXiv:0809.2137 [hep-th].

\bibitem{Hosomichi:2008ip}
  K.~Hosomichi, K.~M.~Lee, S.~Lee, S.~Lee, J.~Park and P.~Yi,
  arXiv:0809.1771 [hep-th].

\bibitem{Yamazaki:2008gg}
  M.~Yamazaki,
  arXiv:0809.1650 [hep-th].

\bibitem{Hanany:2008fj}
  A.~Hanany, D.~Vegh and A.~Zaffaroni,
  arXiv:0809.1440 [hep-th].

\bibitem{deMedeiros:2008zh}
  P.~de Medeiros, J.~Figueroa-O'Farrill, E.~Mendez-Escobar and P.~Ritter,
  arXiv:0809.1086 [hep-th].

\bibitem{Garousi:2008xn}
  M.~R.~Garousi,
  arXiv:0809.0985 [hep-th].

\bibitem{Jeon:2008zj}
  I.~Jeon, J.~Kim, N.~Kim, B.~H.~Lee and J.~H.~Park,
  arXiv:0809.0856 [hep-th].

\bibitem{Garousi:2008yv}
  M.~R.~Garousi,
  arXiv:0809.0381 [hep-th].

\bibitem{Imamura:2008qs}
  Y.~Imamura and K.~Kimura,
  arXiv:0808.4155 [hep-th].

\bibitem{Ueda:2008hx}
  K.~Ueda and M.~Yamazaki,
  arXiv:0808.3768 [hep-th].

\bibitem{Bandos:2008df}
  I.~A.~Bandos,
  arXiv:0808.3568 [hep-th].

\bibitem{Rashkov:2008rm}
  R.~C.~Rashkov,
  arXiv:0808.3057 [hep-th].

\bibitem{Niarchos:2008jb}
  V.~Niarchos,
  arXiv:0808.2771 [hep-th].

\bibitem{Nishioka:2008ib}
  T.~Nishioka and T.~Takayanagi,
  arXiv:0808.2691 [hep-th].

\bibitem{Berenstein:2008dc}
  D.~Berenstein and D.~Trancanelli,
  arXiv:0808.2503 [hep-th].

\bibitem{Iengo:2008cq}
  R.~Iengo and J.~G.~Russo,
  arXiv:0808.2473 [hep-th].

\bibitem{D'Auria:2008cw}
  R.~D'Auria, P.~Fre, P.~A.~Grassi and M.~Trigiante,
  arXiv:0808.1282 [hep-th].

\bibitem{Imeroni:2008cr}
  E.~Imeroni,
  arXiv:0808.1271 [hep-th].

\bibitem{Hanany:2008cd}
  A.~Hanany and A.~Zaffaroni,
  arXiv:0808.1244 [hep-th].

\bibitem{Bonelli:2008us}
  G.~Bonelli, P.~A.~Grassi and H.~Safaai,
  arXiv:0808.1051 [hep-th].

\bibitem{Martelli:2008si}
  D.~Martelli and J.~Sparks,
  arXiv:0808.0912 [hep-th].

\bibitem{Martelli:2008rt}
  D.~Martelli and J.~Sparks,
  arXiv:0808.0904 [hep-th].

\bibitem{Jafferis:2008qz}
  D.~L.~Jafferis and A.~Tomasiello,
  arXiv:0808.0864 [hep-th].

\bibitem{Garousi:2008ai}
  M.~R.~Garousi and A.~Ghodsi,
  arXiv:0808.0411 [hep-th].

\bibitem{Giveon:2008zn}
  A.~Giveon and D.~Kutasov,
  arXiv:0808.0360 [hep-th].

\bibitem{Bak:2008vd}
  D.~Bak, D.~Gang and S.~J.~Rey,
  arXiv:0808.0170 [hep-th].

\bibitem{Nilsson:2008kq}
  B.~E.~W.~Nilsson and J.~Palmkvist,
  arXiv:0807.5134 [hep-th].

\bibitem{Bonelli:2008kh}
  G.~Bonelli, A.~Tanzini and M.~Zabzine,
  arXiv:0807.5113 [hep-th].

\bibitem{Singh:2008ix}
  H.~Singh,
  arXiv:0807.5016 [hep-th].

\bibitem{Aharony:2008gk}
  O.~Aharony, O.~Bergman and D.~L.~Jafferis,
  arXiv:0807.4924 [hep-th].

\bibitem{Gromov:2008fy}
  N.~Gromov and V.~Mikhaylov,
  arXiv:0807.4897 [hep-th].

\bibitem{Krishnan:2008zs}
  C.~Krishnan,
  arXiv:0807.4561 [hep-th].

\bibitem{Alday:2008ut}
  L.~F.~Alday, G.~Arutyunov and D.~Bykov,
  arXiv:0807.4400 [hep-th].

\bibitem{Kluson:2008nw}
  J.~Kluson,
  arXiv:0807.4054 [hep-th].

\bibitem{McLoughlin:2008ms}
  T.~McLoughlin and R.~Roiban,
  arXiv:0807.3965 [hep-th].

\bibitem{Honma:2008ef}
  Y.~Honma, S.~Iso, Y.~Sumitomo, H.~Umetsu and S.~Zhang,
  arXiv:0807.3825 [hep-th].

\bibitem{Ahn:2008hj}
  Changrim Ahn, P.~Bozhilov and R.~C.~Rashkov,
  JHEP {\bf 0809}, 017 (2008).

\bibitem{Shenderovich:2008bs}
  I.~Shenderovich,
  arXiv:0807.2861 [hep-th].

\bibitem{Bergshoeff:2008bh}
  E.~A.~Bergshoeff, O.~Hohm, D.~Roest, H.~Samtleben and E.~Sezgin,
  arXiv:0807.2841 [hep-th].

\bibitem{Lee:2008ui}
  B.~H.~Lee, K.~L.~Panigrahi and C.~Park,
  arXiv:0807.2559 [hep-th].

\bibitem{Imamura:2008dt}
  Y.~Imamura and K.~Kimura,
  arXiv:0807.2144 [hep-th].

\bibitem{Verlinde:2008di}
  H.~Verlinde,
  arXiv:0807.2121 [hep-th].

\bibitem{Hanaki:2008cu}
  K.~Hanaki and H.~Lin,
  arXiv:0807.2074 [hep-th].

\bibitem{Bak:2008cp}
  D.~Bak and S.~J.~Rey,
  arXiv:0807.2063 [hep-th].

\bibitem{Ahn:2008aa}
  Changrim Ahn and R.~I.~Nepomechie,
  arXiv:0807.1924 [hep-th].

\bibitem{Astolfi:2008ji}
  D.~Astolfi, V.~G.~M.~Puletti, G.~Grignani, T.~Harmark and M.~Orselli,
  arXiv:0807.1527 [hep-th].

\bibitem{Hashimoto:2008iv}
  A.~Hashimoto and P.~Ouyang,
  arXiv:0807.1500 [hep-th].

\bibitem{Garousi:2008ik}
  M.~R.~Garousi, A.~Ghodsi and M.~Khosravi,
  JHEP {\bf 0808}, 067 (2008).

\bibitem{Pang:2008hw}
  Y.~Pang and T.~Wang,
  arXiv:0807.1444 [hep-th].

\bibitem{Kim:2008gn}
  N.~Kim,
  arXiv:0807.1349 [hep-th].

\bibitem{Li:2008ya}
  T.~Li, Y.~Liu and D.~Xie,
  arXiv:0807.1183 [hep-th].

\bibitem{Schnabl:2008wj}
  M.~Schnabl and Y.~Tachikawa,
  arXiv:0807.1102 [hep-th].

\bibitem{Zhou:2008sa}
  Y.~Zhou,
  arXiv:0807.0890 [hep-th].

\bibitem{Chu:2008qv}
  C.~S.~Chu, P.~M.~Ho, Y.~Matsuo and S.~Shiba,
  JHEP {\bf 0808}, 076 (2008).

\bibitem{Cherkis:2008qr}
  S.~Cherkis and C.~Saemann,
  arXiv:0807.0808 [hep-th].

\bibitem{Chen:2008qq}
  B.~Chen and J.~B.~Wu,
  arXiv:0807.0802 [hep-th].

\bibitem{Gromov:2008qe}
  N.~Gromov and P.~Vieira,
  arXiv:0807.0777 [hep-th].

\bibitem{Ahn:2008gd}
  Changrim Ahn and P.~Bozhilov,
  JHEP {\bf 0808}, 054 (2008).

\bibitem{Gromov:2008bz}
  N.~Gromov and P.~Vieira,
  arXiv:0807.0437 [hep-th].

\bibitem{Terashima:2008ba}
  S.~Terashima and F.~Yagi,
  arXiv:0807.0368 [hep-th].

\bibitem{Grignani:2008te}
  G.~Grignani, T.~Harmark, M.~Orselli and G.~W.~Semenoff,
  arXiv:0807.0205 [hep-th].

\bibitem{Terashima:2008sy}
  S.~Terashima,
  JHEP {\bf 0808}, 080 (2008).

\bibitem{Bagger:2008se}
  J.~Bagger and N.~Lambert,
  arXiv:0807.0163 [hep-th].

\bibitem{Beisert:2008qy}
  N.~Beisert,
  arXiv:0807.0099 [hep-th].

\bibitem{Okuyama:2008qd}
  K.~Okuyama,
  arXiv:0807.0047 [hep-th].

\bibitem{Fre:2008qc}
  P.~Fre and P.~A.~Grassi,
  arXiv:0807.0044 [hep-th].

\bibitem{Hosomichi:2008jb}
  K.~Hosomichi, K.~M.~Lee, S.~Lee, S.~Lee and J.~Park,
  arXiv:0806.4977 [hep-th].

\bibitem{Grignani:2008is}
  G.~Grignani, T.~Harmark and M.~Orselli,
  arXiv:0806.4959 [hep-th].

\bibitem{Stefanski:2008ik}
  B.~j.~Stefanski,
  arXiv:0806.4948 [hep-th].

\bibitem{Arutyunov:2008if}
  G.~Arutyunov and S.~Frolov,
  arXiv:0806.4940 [hep-th].

\bibitem{Bedford:2008hn}
  J.~Bedford and D.~Berman,
  arXiv:0806.4900 [hep-th].

\bibitem{Gaiotto:2008cg}
  D.~Gaiotto, S.~Giombi and X.~Yin,
  arXiv:0806.4589 [hep-th].

\bibitem{Agarwal:2008rr}
  A.~Agarwal,
  arXiv:0806.4292 [hep-th].

\bibitem{Hanany:2008qc}
  A.~Hanany, N.~Mekareeya and A.~Zaffaroni,
  arXiv:0806.4212 [hep-th].

\bibitem{Armoni:2008kr}
  A.~Armoni and A.~Naqvi,
  arXiv:0806.4068 [hep-th].

\bibitem{Minahan:2008hf}
  J.~A.~Minahan and K.~Zarembo,
  arXiv:0806.3951 [hep-th].

\bibitem{Imamura:2008nn}
  Y.~Imamura and K.~Kimura,
  arXiv:0806.3727 [hep-th].

\bibitem{Honma:2008jd}
  Y.~Honma, S.~Iso, Y.~Sumitomo and S.~Zhang,
  arXiv:0806.3498 [hep-th].

\bibitem{Nishioka:2008gz}
  T.~Nishioka and T.~Takayanagi,
  JHEP {\bf 0808}, 001 (2008).

\bibitem{Blau:2008bm}
  M.~Blau and M.~O'Loughlin,
  arXiv:0806.3253 [hep-th].

\bibitem{Bhattacharya:2008bja}
  J.~Bhattacharya and S.~Minwalla,
  arXiv:0806.3251 [hep-th].

\bibitem{deMedeiros:2008bf}
  P.~de Medeiros, J.~M.~Figueroa-O'Farrill and E.~Mendez-Escobar,
  JHEP {\bf 0808}, 045 (2008).

\bibitem{Bergshoeff:2008ix}
  E.~A.~Bergshoeff, M.~de Roo, O.~Hohm and D.~Roest,
  arXiv:0806.2584 [hep-th].

\bibitem{Mauri:2008ai}
  A.~Mauri and A.~C.~Petkou,
  arXiv:0806.2270 [hep-th].

\bibitem{Cecotti:2008qs}
  S.~Cecotti and A.~Sen,
  arXiv:0806.1990 [hep-th].

\bibitem{Ezhuthachan:2008ch}
  B.~Ezhuthachan, S.~Mukhi and C.~Papageorgakis,
  JHEP {\bf 0807}, 041 (2008).

\bibitem{Benna:2008zy}
  M.~Benna, I.~Klebanov, T.~Klose and M.~Smedback,
  arXiv:0806.1519 [hep-th].

\bibitem{BLS}
  M.~A.~Bandres, A.~E.~Lipstein and J.~H.~Schwarz,
  JHEP {\bf 0807}, 117 (2008).


\end{thebibliography}
\end{document}